\documentclass[pdflatex,sn-mathphys-num, iicol]{sn-jnl}

\usepackage{graphicx}%
\usepackage{multirow}%
\usepackage{amsmath,amssymb,amsfonts}%
\usepackage{amsthm}%
\usepackage{mathrsfs}%
\usepackage[title]{appendix}%
\usepackage{xcolor}%
\usepackage{textcomp}%
\usepackage{manyfoot}%
\usepackage{booktabs}%
\usepackage{algorithm}%
\usepackage{algorithmicx}%
\usepackage{algpseudocode}%
\usepackage{listings}%
\usepackage{amsmath}
\PassOptionsToPackage{hyphens}{url} 
\usepackage{hyperref}               
\hypersetup{
  breaklinks=true,
  colorlinks=true,
  urlcolor=blue
}
\usepackage[numbers]{natbib}

\theoremstyle{thmstyleone}%
\theoremstyle{thmstyletwo}%

\theoremstyle{thmstylethree}%

\begin{document}

\title{\textbf{Study of anisotropic flow of heavy hadrons in Au + Au collisions at $\sqrt{s_{NN}} =$ 200 GeV using HYDJET++ framework}}


\author[1]{\fnm{Sunidhi} \sur{Saxena}}

\author[1]{\fnm{Gauri} \sur{Devi}}

\author[1]{\fnm{Rajiv} \sur{Gupta}}

\author*[1]{\fnm{Ajay} \sur{Kumar}}\email{ajay.phy@bhu.ac.in}

\affil[1]{\orgdiv{Department of Physics}, \orgname{Institute of Science}, \orgaddress{\street{Banaras Hindu University (BHU)}, \city{Varanasi}, \postcode{221005}, \country{India}}}


\abstract{A comprehensive study of the anisotropic flow of heavy hadrons ($D^{0}$, $D^{\pm}$, and $\Lambda_{c}$) in Au + Au collisions at $\sqrt{s_{NN}} = 200$ GeV using the HYDJET++ model is presented. This study aims to explore the collective behavior and thermalization of charm hadrons at RHIC energy. The modeling of anisotropic flow is performed using a centrality-dependent parameterization of anisotropic parameters. Our model results are consistent with STAR experimental results up to $p_{T} = 4$ GeV/c. It captures the centrality-dependent shifts of the flow peak towards lower $p_{T}$ as collisions become more peripheral. This shift is attributed to the weaker radial flow in peripheral collisions. The model also reproduces important features such as the number-of-constituent-quark scaling, mass ordering, and baryon–meson grouping, all consistent with experimental observations. Furthermore, we present comparisons with other models, like DUKE, SUBATECH, TAMU, and AMPT. Overall, our results highlight the efficacy of HYDJET++ in describing the collective dynamics of heavy-flavor hadrons in the quark-gluon plasma medium.}




\maketitle

\section{Introduction}\label{sec:intro}

Under extreme conditions, such as high energy density ($\sim 1\,\mathrm{GeV}/\mathrm{fm}^{3}$) and high temperature ($\sim 170 MeV$) \citep{PHENIX:2004vcz}, the confined hadronic state transitions into a phase of deconfined quarks and gluons. This new state of Quantum Chromodynamic (QCD) matter is known as the Quark-Gluon Plasma (QGP). The QGP is created at the Relativistic Heavy Ion Collider (RHIC) and the Large Hadron Collider (LHC) through collisions of heavy ions at ultra-relativistic speeds. During these collisions, the QGP constituents exhibit significant interactions, leading to collective, hydrodynamic, fluid-like behavior. Hence, their collective behavior is modeled by hydrodynamic models and can be studied in a high-energy collision through the uneven distribution of momentum among particles (light and heavy hadrons), referred to as anisotropic flow \cite{Gale:2012rq}. The present study focuses on heavy hadrons, which are composed of heavy quarks produced during the early stages of nucleus-nucleus collisions \cite{Cacciari:2005rk}. These heavy quarks undergo Brownian-like motion throughout the entire evolution of the QGP \cite{Rapp:2008qc}. Therefore, the study of the anisotropy of heavy hadrons provides information about the QGP medium without being affected by the production mechanisms of the heavy quarks.

The anisotropic flow resulting from multi-parton production serves as a valuable probe for studying the collective behavior of the QGP \cite{Ollitrault:1992bk, Voloshin:2008dg}. In non-central nucleus-nucleus collisions, the overlap region between the colliding nuclei is not spatially isotropic. Instead, it takes on an almond-shaped (elliptic) geometry in the transverse plane relative to the reaction plane. 
Due to parton re-interactions, a pressure gradient develops, which is weaker along the major axis of the ellipse. This leads to weaker flow along the major axis and stronger flow along the minor axis. Hence, the spatial anisotropy pushes more particles along the minor axis than the major axis, resulting in momentum anisotropy. These momentum anisotropies can be described using a Fourier series expansion \cite{Voloshin:1994mz, Poskanzer:1998yz, Voloshin:2008dg}:
\begin{equation}
\frac{dN}{d\phi} = 1 + 2\sum_{n=1}^\infty v_{n} cos[n(\phi - \psi_{n})]    
\end{equation}
Here, $\phi$ is the azimuthal angle between the momentum of the produced particle and the reaction plane $\psi_{n}$; $n$ is the harmonic number, and $v_{n}$ represents the $n^{th}$-order harmonic coefficient of anisotropic flow, which can be expressed as:
\begin{equation}
    v_{n} = \langle \langle \cos[n(\phi - \psi_{n})] \rangle \rangle
\end{equation}
Here, $v_{1}$ is directed flow, $v_{2}$ is elliptic flow, $v_{3}$ is triangular flow, $v_{4}$ is quadrangular flow, and so on. Directed flow ($v_{1}$) \cite{Zabrodin:2000xc} helps study the expansion of the QGP. Since $v_{1}$ develops over time, it reflects the conditions during the later stages of the collision. The measurement of $v_{1}$ at high transverse momentum ($p_{T}$) provides a sensitive probe of the early, high-density stage of the system, offering insights into the Equation of State (EOS) and may offer indications of a first-order phase transition in the QCD phase diagram. Elliptic flow ($v_{2}$) \cite{Voloshin:2008dg} reflects the pressure developed during the initial stage of QGP formation and indicates its thermalization timescale. Among the flow harmonics, the contribution of $v_{2}$ is the largest. Triangular flow $v_{3}$ \cite{Voloshin:2008dg} arises due to initial state fluctuations in the positions of participating nucleons at the time of collision. The quadrangular flow $v_{4}$ \cite{Kolb:2003zi} arises due to initial eccentricities and hydrodynamic expansion of QGP. In this study, we focus on flow harmonics up to $v_{4}$, as higher-order components are primarily driven by fluctuations and are less directly connected to the fundamental properties of the QGP medium.

The first measurement of the anisotropic flow of the $D^{0}$ meson at the STAR experiment in Au + Au collisions at center-of-mass energy ($\sqrt{s_{NN}}$) of 200 GeV at mid-rapidity ($|y| < 1$) was performed using the 2014 dataset (Run 2014) \cite{STAR:2017kkh}. This observation showed that at low $p_{T}$, the elliptic flow of the $D^{0}$ meson is smaller than that of light hadrons, indicating the presence of mass ordering. At high $p_{T}$, the $D^{0}$ meson exhibits collective flow similar to light mesons. Furthermore, the study demonstrated that the $D^{0}$ meson also follows the number of constituent quark (NCQ) scaling, similar to light hadrons, which suggests that heavy mesons are fully thermalized with the medium.
In addition, data from the 2016 run, which benefited from an improved detector system, was also analyzed \cite{Michael, Liang}. Following this, the STAR collaboration presented combined results from Run 2014 and Run 2016 \cite{Liang}. This analysis includes the first measurement of the triangular flow of $D^{0}$ mesons across various centrality bins. The non-zero value of $v_{3}$ highlights the significance of initial-state fluctuations in the collisions. Additionally, preliminary results of $v_{2}$ for $D^{\pm}$ meson in the 0-80\% and 10-40\% centrality bins at $p_{T} > 2$ GeV/c were also reported. 

The study of the anisotropic flow of heavy hadrons has also been performed using various models such as   DUKE \cite{Cao:2015hia}, SUBATECH \cite{Ozvenchuk:2014rpa, Nahrgang:2014vza}, TAMU \cite{He:2011qa}, AMPT \cite{STAR:2017kkh}, and HYDJET++ \cite{Lokhtin:2008xi}.
The DUKE group \cite{Cao:2015hia} uses the MC-Glauber model for spatial distribution, leading-order perturbative QCD for the momentum space distribution \cite{Combridge:1978kx}, CTEQ parton distribution functions (PDFs) \cite{Lai:1999wy}, and the EPS09 parameterization \cite{Eskola:2009uj} for shadowing/anti-shadowing effects. The QGP medium is simulated using (2+1)-dimensional viscous hydrodynamics \cite{Qiu:2011hf}, and heavy quark evolution is modeled via a modified Langevin equation  \cite{Cao:2013ita}. Hadronization occurs through fragmentation and recombination, followed by UrQMD-modeled hadronic interactions \cite{Bass:1998ca}. This framework describes the nuclear modification factor ($R_{AA}$) and elliptic flow ($v_{2}$) at low $p_{T}$. The SUBATECH group \cite{Ozvenchuk:2014rpa, Nahrgang:2014vza} employs MC-Glauber and FONLL for initial conditions \cite{Cacciari:1998it, Cacciari:2001td, Cacciari:2012ny} with CTEQ6.6 PDFs \cite{Nadolsky:2008zw}. The QGP medium is modeled using 3+1 dimensional ideal fluid dynamics \cite{Werner:2010aa}, with heavy quark evolution governed by the Boltzmann equation \cite{Gossiaux:2010yx}, hadronization described via fragmentation and coalescence mechanisms, and hadronic interactions modeled using the Fokker-Planck equation \cite{Ozvenchuk:2014rpa}. This framework successfully describes the STAR experimental data \cite{STAR:2017kkh} for $R_{AA}$ and $v_{2}$ up to $p_{T}=2$ GeV/c. The TAMU group \cite{He:2011qa} applies non-perturbative T-matrix approaches \cite{Riek:2010fk, Riek:2010py} to determine the transport properties and Fokker-Planck-Langevin dynamics to describe the diffusion of heavy quarks. The QGP medium is modeled by the AZHYDRO code \cite{Kolb:2000sd, Kolb:2003dz}, which is based on 2+1-dimensional hydrodynamics. Hadronization via coalescence is treated using the resonance recombination model \cite{Ravagli:2007xx}. This model framework with charm quark diffusion, describes the STAR experimental data \cite{STAR:2017kkh} up to $p_{T}=2$ GeV/c. A multi-phase transport (AMPT) model \cite{Lin:2004en} incorporates HIJING event generators for initial conditions \cite{Gyulassy:1994ew}, ZPC model \cite{Zhang:1997ej} for partonic interactions, Lund string fragmentation model \cite{Andersson:1997xwk} for fragmentation and improved string melting version of AMPT (AMPT-SM) \cite{Lin:2001zk} for the quark coalescence mechanism. The ART hadronic transport model \cite{Li:2001xh} describes the hadronic interactions. This framework effectively describes the STAR experimental results of $v_{2}$ for transverse momentum up to $p_{T}<4$ GeV/c \cite{STAR:2017kkh}.
 
The HYDJET++ model adopts a two-component approach, consisting of soft and hard states \cite{Lokhtin:2008xi}. In this framework, the soft thermal component is simulated using the FAST MC generator \cite{Amelin:2006qe, Amelin:2007ic}, while the hard component is modeled with the PYQUEN event generator, which incorporates parton energy loss \cite{Lokhtin:2005px}. The medium is represented as a boost-invariant, longitudinally expanding quark-gluon fluid. Final-state hadrons and resonances are sampled from the SHARE particle data table \cite{Torrieri:2004zz}. The HYDJET++ model has successfully reproduced anisotropic flow across various collision systems at RHIC and LHC energies \cite{Singh:2017fgm, Pandey:2021xjn, Devi:2023wih}. It also describes the anisotropic flow of charm hadrons in Pb+Pb collisions at the LHC energy of $\sqrt{s_{NN}}=2.76$ TeV \cite{Lokhtin:2017rvj}. However, at RHIC energies, the study of heavy-hadron flow remains largely unexplored. In this work, we investigate the anisotropic flow of charm hadrons in Au + Au collisions at a center-of-mass energy of $\sqrt{s_{NN}}=200$ GeV using the HYDJET++ model. This study enhances our understanding of the collective behavior of charm hadrons within the QGP medium and supports our previous findings \cite{Saxena:2025sys} on the thermalization of charm hadrons in such an environment.

This paper is organized as follows: Section \ref{sec:model} details the implementation of anisotropic flow in the HYDJET++ model. Section \ref{sec:results} presents charm hadron flow results as a function of transverse momentum, comparing HYDJET++ predictions with experimental data and other models, including higher-order flow harmonics. Section \ref{sec:summ} summarizes the study and outlines future directions.

\section{Description of anisotropic flow in HYDJET++ model}\label{sec:model}

HYDJET++ (an upgraded version of the HYDJET event generator \cite{Lokhtin:2005px}) simulates heavy-ion collisions by combining two distinct components: a soft, hydrodynamic-like "thermal" state and a hard or "non-thermal" state, resulting from medium-modified multi-parton fragmentation. In this framework, the soft and hard components are generated independently. Detailed information about the physics model and simulation procedure is available in the HYDJET++ manual \cite{Lokhtin:2008xi, Bravina:2013xla}. In brief, the soft component represents a thermal hadronic state that forms at chemical and thermal freeze-out hypersurfaces. It is modeled using a parameterized relativistic hydrodynamic framework with fixed freeze-out conditions and is generated through the FASTMC event generator \cite{Amelin:2006qe, Amelin:2007ic}. For the thermal production of charm hadrons, HYDJET++ adopts a statistical hadronization approach \cite{Andronic:2003zv, Andronic:2006ky}. Hadron multiplicities are estimated using the effective thermal volume approximation. Additionally, the stochastic nature of hadron production is modeled using the Poisson distribution centered around a mean value and proportional to the number of participating nucleons at a given impact parameter in an AA collision \cite{Bravina:2013xla}. To incorporate the effects of elliptic flow, a hydro-inspired parametrization is applied to describe the momentum and spatial anisotropy of the soft hadron emission source \cite{Lokhtin:2008xi, Wiedemann:1997cr}.

The hard component of the HYDJET++ model is identical to that of the HYDJET model \cite{Lokhtin:2005px, Lokhtin:2007ga}. It is simulated using the PYTHIA \cite{Sjostrand:2006za} and PYTHIA QUENCHED (PYQUEN) \cite{Lokhtin:2005px} event generators. PYTHIA is used to produce the initial parton spectra from hard nucleon-nucleon (NN) scatterings, while the PYQUEN model simulates partonic energy loss, incorporating impact parameter-dependent binary nucleon collision vertices derived using the Glauber model \cite{Loizides:2017ack}. The number of PYQUEN jets in an event follows a binomial distribution. The mean number of jets is calculated as the product of the number of binary nucleon-nucleon (NN) sub-collisions at a given impact parameter and the integral cross-section of the hard process in NN collisions. This calculation considers only processes above a minimum transverse momentum transfer ($p_{T}^{min}$), which is an input parameter of the model. In HYDJET++, partons produced in (semi)hard processes with momentum transfer below $p_{T}^{min}$ are assumed to be "thermalized," and their hadronization products are automatically included in the soft component of the event. The approach to describing multiple scattering of hard partons focuses on their energy loss. This energy loss accumulates through gluon radiation at each scattering event as the partons propagate through the expanding quark-gluon plasma. It incorporates the interference effects in gluon emission with a finite formation time, using a modified radiation spectrum $dE/dx$ that varies with the decreasing temperature $T$. The model considers both radiative and collisional energy losses of hard partons within a longitudinally expanding quark-gluon medium and incorporates realistic nuclear geometry. To account for nuclear shadowing effects on parton distribution functions, an impact parameter-dependent parametrization \cite{Tywoniuk:2007xy} based on the Glauber–Gribov theory is employed. 

In contrast to full-scale relativistic hydrodynamic models (which require significant computational resources), HYDJET++ employs simple and commonly used parametrizations of the freeze-out hypersurface \cite{Lokhtin:2008xi} to simulate higher-order azimuthal anisotropy harmonics. As a result, the anisotropic elliptic shape of the initial nuclear overlap translates into a corresponding anisotropy in the final momentum distribution of the emitted particles.
To describe the second harmonic $v_{2}$, the model employs two coefficients, $\delta(b)$ and $\epsilon(b)$, which represent the flow anisotropy and the spatial (coordinate) anisotropy of the fireball at freeze-out, respectively, both as functions of the impact parameter (b). These parameters can either be adjusted independently for each centrality class or linked through their dependence on the initial geometric ellipticity $\epsilon_{0}(b)=b/2R_{A}$, where $R_{A}$ is the nuclear radius. In hydrodynamic approach \cite{Wiedemann:1997cr} (used in HYDJET++), elliptic flow can be defined as \cite{Lokhtin:2008xi}:

\begin{equation}
   v_{2} \propto \frac{2(\delta - \epsilon)}{(1-\delta^{2})(1-\epsilon^{2})} 
\end{equation}

Furthermore, the non-elliptic geometry of the initial overlap region in nuclear collisions is quantified by the initial triangular coefficient $\epsilon_{3}(b)$, which leads to the emergence of higher-order Fourier harmonics in the final momentum distribution. HYDJET++ model enables a straightforward parametrization of this anisotropy through the natural modulation of the freeze-out hypersurface, which can be described by the equation below \cite{Bravina:2013xla}:
\begin{align}
R(b,\phi) &= R_{f}(b) \frac{\sqrt{1 - \epsilon^{2}(b)}}{\sqrt{1 + \epsilon(b)\cos 2\phi}} \nonumber \\
         &\quad \times \left[1 + \epsilon_{3}(b)\cos 3(\phi + \psi_{3}^{RP})\right]
\end{align}

Here, $\phi$ denotes the spatial azimuthal angle of a fluid element relative to the impact parameter direction. The transverse radius of the fireball in the azimuthal direction $\phi$ is given by $R(b,\phi)$, scaled by a model parameter $R_{f}(b)$. The phase $\psi_{3}^{RP}$ introduces the third harmonic component, which has a randomly oriented reaction plane with respect to the impact parameter direction (where $\psi_{2}^{RP}=0$). The new anisotropy parameter $\epsilon_{3}(b)$ can be treated independently for each centrality class or expressed in terms of the initial ellipticity $\epsilon_{0}(b)$. Significantly, this modulation does not influence the elliptic flow coefficient $v_{2}$, which was previously fitted using the parameters $\delta(b)$ and $\epsilon(b)$ \cite{Lokhtin:2008xi}.

\begin{figure*}[tbp]
\centering
\includegraphics[width=15cm]{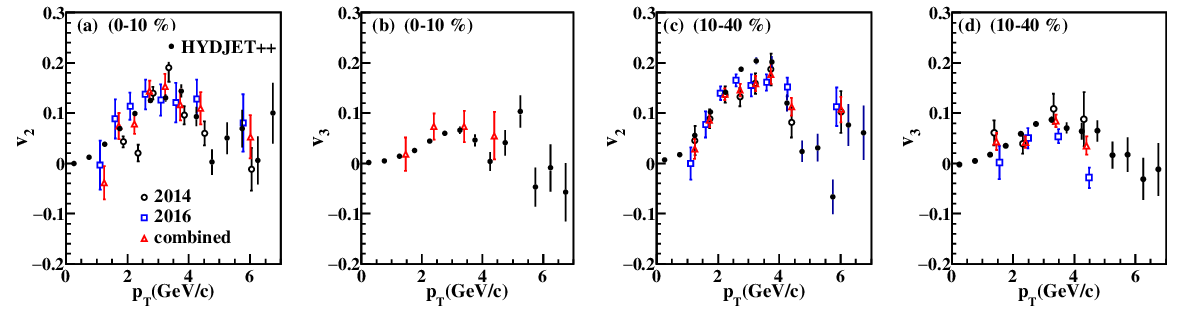}
\caption{\label{figure1}Elliptic flow ($v_{2}$) and triangular flow ($v_{3}$) of $D^{0}$ meson as a function of transverse momentum ($p_{T}$) for 0-10\% and 10-40\% centrality bins in Au + Au collisions at $\sqrt{s_{\mathrm{NN}}} = 200$ GeV. Open markers represent three datasets from the STAR experiment \cite{STAR:2017kkh, Michael, Liang}, while closed markers show results from the HYDJET++ model.}

\end{figure*}

\begin{figure*}[tbp]
\centering
\includegraphics[width=15cm]{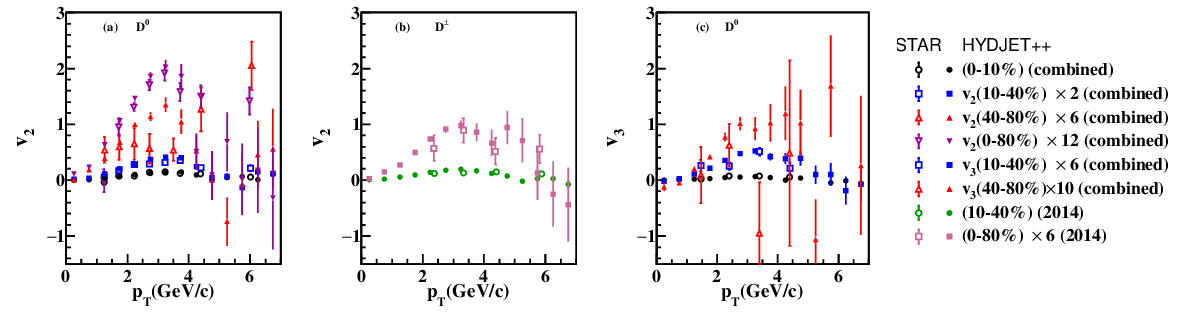}
\caption{\label{figure2}Elliptic flow ($v_{2}$) of $D^{0}$ and $D^{\pm}$ mesons, and triangular flow ($v_{3}$) of $D^{0}$ mesons as a function of transverse momentum ($p_{T}$) for various centrality classes in Au + Au collisions at $\sqrt{s_{\mathrm{NN}}} = 200$ GeV. Open markers represent data from STAR measurements \cite{STAR:2017kkh, Michael, Liang}, while closed markers represent results from the HYDJET++ model.}

\end{figure*}
All produced particles originate from a thermally expanding source characterized by a maximum transverse flow velocity ($\rho_{u}$) \cite{Lokhtin:2008xi}, which is defined as:
\begin{equation}
    \rho_{u}^{max}=\rho_{u}^{max}(b=0)[1+\rho_{3u}(b)cos3\phi+\rho_{4u}(b)cos4\phi]
\end{equation}

This parametrization of $\rho_{u}$ also allows the inclusion of higher-order azimuthal harmonics, though they remain aligned with the impact parameter direction $(\psi_{2}^{RP}=0)$. In this scenario, the velocity profile becomes modulated across the entire freeze-out hypersurface, and this modulation cannot be rotated independently using a separate phase. The newly introduced anisotropy parameters, $\rho_{3u}(b)$ and $\rho_{4u}(b)$ can again be handled independently for each centrality class or expressed in terms of the initial ellipticity $\epsilon_{0}(b)$.

In this study, we employ a centrality-dependent parameterization of the anisotropy parameters to model the anisotropic flow. This study is in the continuation of our earlier work \cite{Saxena:2025sys} on heavy hadrons production in Au + Au collisions at 200 GeV RHIC energy, which is consistent with the STAR experimental data \cite{STAR:2018zdy, Vanek:2022ekr, STAR:2019ank}.

\section{Results}
\label{sec:results}
\subsection{Comparison of HYDJET++ results with experimental data}

\subsubsection{$p_{T}$ dependent $v_{n}$ of heavy hadrons}

\begin{figure*}[tbp]
\centering
\includegraphics[width=12cm]{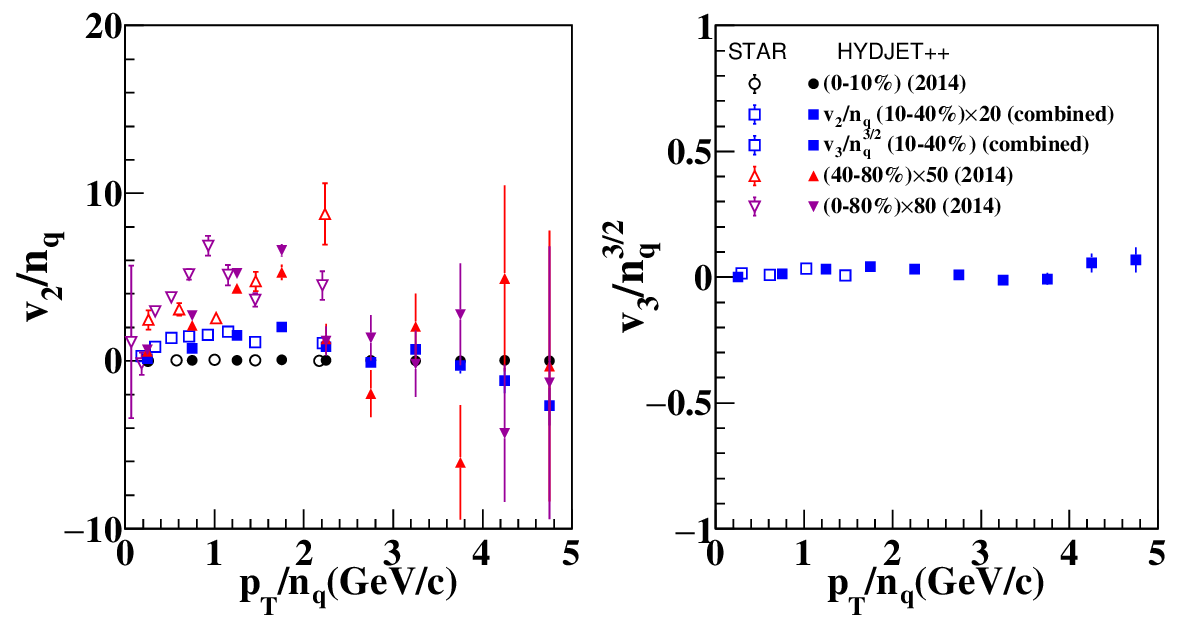}
\caption{\label{figure3}NCQ scaling of $D^{0}$ mesons, shown as $v_{2}/n_{q}$ versus $p_{T}/n_{q}$ and $v_{3}/n_{q}^{3/2}$ versus $p_{T}/n_{q}$ for different centrality classes in Au + Au collisions at $\sqrt{s_{\mathrm{NN}}} = 200$ GeV. Open markers represent data from STAR experiment \cite{STAR:2017kkh, Michael, Liang}, while closed markers represent results from the HYDJET++ model.}

\end{figure*}
Figure \ref{figure1} represents the elliptic flow ($v_{2}$) and triangular flow ($v_{3}$) of $D^{0}$ mesons as a function of $p_{T}$ in Au + Au collisions at $\sqrt{s_{NN}}=200$ GeV for the 0-10\% and 10-40\% centrality intervals. The HYDJET++ model results are compared with three types of datasets from the STAR experiment: Run 2014, Run 2016, and the combined dataset \cite{STAR:2017kkh, Michael, Liang}. At low transverse momentum ($p_{T} \leq 3$ GeV/c), both $v_{2}$ and $v_{3}$ exhibit a linear dependence on $p_{T}$ due to collisional energy loss. At higher $p_{T}$ ($\geq3$ GeV/c), both $v_{2}$ and $v_{3}$ decrease with increasing $p_{T}$ due to radiative energy loss. A peak is also observed at $p_{T}\approx 3$ GeV/c, attributed to bulk flow effects \cite{Hong:2025dfj}. A similar trend is observed in the experimental data \cite{STAR:2017kkh, Michael, Liang}. It is observed that our model results are consistent with the combined dataset; therefore, in further analysis, we will compare our model predictions only with the combined dataset \cite{Liang}.

Figure \ref{figure2} represents the $v_{2}(p_{T})$ of $D^{0}$ and $D^{\pm}$ mesons, as well as the $v_{3}(p_{T})$ of $D^{0}$ mesons, for various centrality intervals. It is observed that as we move from central to peripheral collisions, the peak in $v_{2}$ shifts towards lower $p_{T}$. This behavior is attributed to the weaker radial flow during the QGP evolution in peripheral collisions compared to central ones \cite{Hong:2025dfj}. Additionally, the magnitude of flow increases as we move from central to peripheral collision. This is due to the large eccentricity in peripheral collision compared to central collision \cite{Hong:2025dfj}. Our model results match with the experimental data from the STAR experiment \cite{Michael, Liang} up to $p_{T}=4$ GeV/c for the 0-10\% and 10-40\% centrality intervals. At higher $p_{T} ~(> 4$ GeV/c), the model underpredicts the experimental data due to the presence of non-flow effects from jets  \cite{Bravina:2013xla}. For the 40–80\% centrality interval, our model results match the experimental data \cite{Liang} up to $p_{T}=2.5$ GeV/c; beyond this $p_{T}$ value ($p_{T}>2.5$ GeV/c), the overprediction can be improved by incorporating mini-jet production or other similar mechanisms \cite{Bravina:2013xla}.  
\subsubsection{Number-of-Constituent-Quark (NCQ) scaling in heavy hadron flow}

Figure \ref{figure3} represents the NCQ scaling, studied through $v_{2}/n_{q}$ as a function of $p_{T}/n_{q}$ and $v_{3}/n_{q}^{3/2}$ as a function of $p_{T}/n_{q}$, where $n_{q}$ is the number of constituents quark in a hadron. Our model results for $v_{2}/n_{q}$ indicate that $D^{0}$ mesons follow NCQ scaling across the entire $p_{T}/n_{q}$ range and are consistent with the experimental results \cite{STAR:2017kkh, Michael, Liang} for the 0–10\% and 10–40\% centrality classes. For the 40-80\% and 0-80\% centrality intervals, the magnitude of $v_{2}/n_{q}$ from our model is slightly lower than the experimental data \cite{STAR:2017kkh, Michael}. Additionally, the NCQ scaling of the triangular flow, i.e., $v_{3}/n_{q}^{3/2}$, matches the experimental results \cite{Liang}. 
An exact match between the model and experiment is optional for validating NCQ scaling. Hence, we proceeded with predicting NCQ scaling in Section \ref{section3}.

\subsection{Comparison of HYDJET++ results with other models}

\begin{figure*}[tbp]
\centering
\includegraphics[width=12cm]{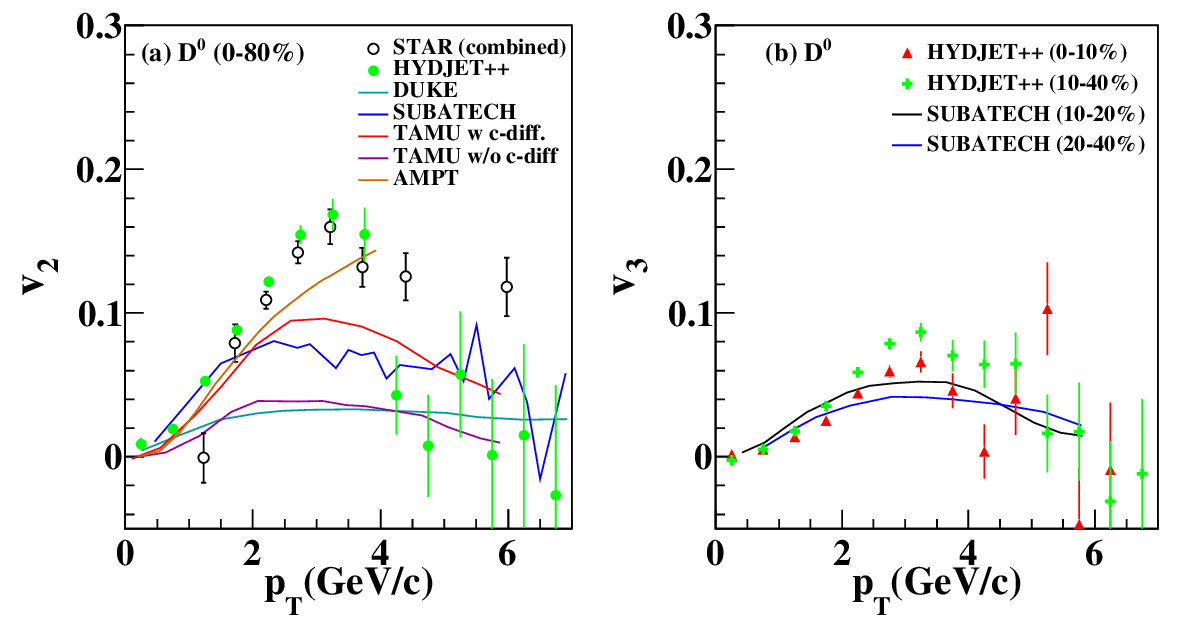}
\caption{\label{figure4} Elliptic flow ($v_{2}$) and triangular flow ($v_{3}$) of $D^{0}$ mesons as a function of transverse momentum ($p_{T}$) for different centrality bins in Au + Au collisions at $\sqrt{s_{\mathrm{NN}}} = 200$ GeV. Solid lines represent various model calculations, including DUKE \cite{Cao:2015hia}, SUBATECH \cite{Ozvenchuk:2014rpa, Nahrgang:2014vza}, TAMU with charm diffusion \cite{He:2011qa}, TAMU without charm diffusion \cite{He:2011qa}, and AMPT \cite{Lin:2004en}. Open markers represent results from the STAR experiment \cite{Liang}, while closed markers represent HYDJET++ model results.}

\end{figure*}

\begin{figure*}[tbp]
\centering
\includegraphics[width=12cm]{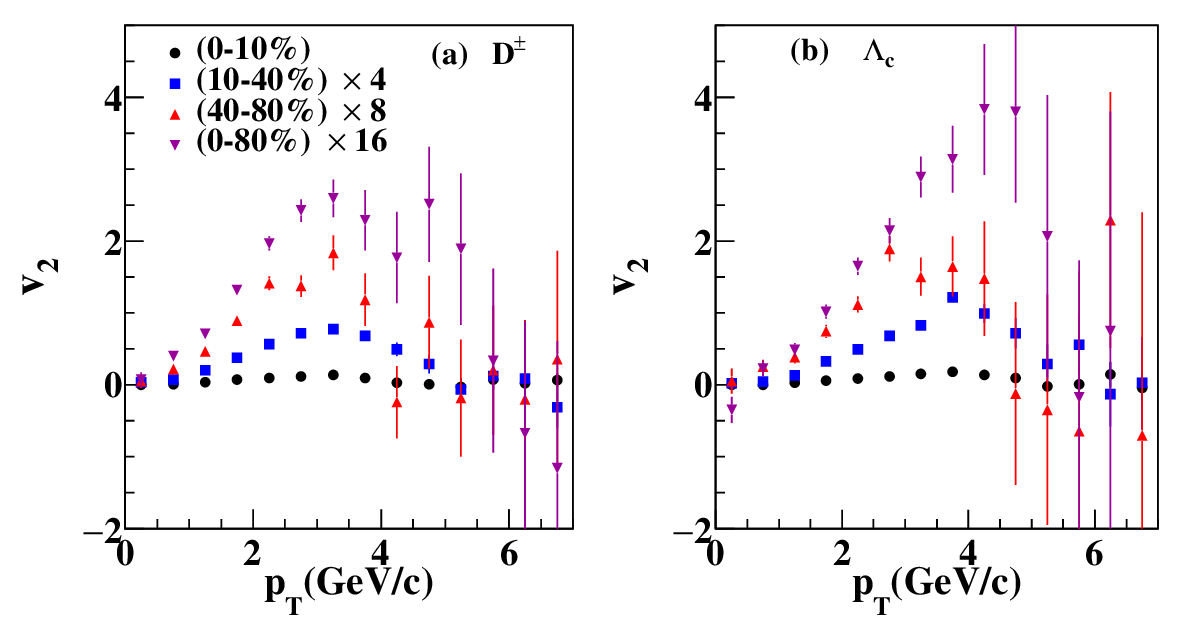}
\caption{\label{figure5} Elliptic flow ($v_{2}$) of $D^{\pm}$ mesons and $\Lambda_{c}$ baryons as a function of transverse momentum ($p_{T}$) for different centrality classes in Au + Au collisions at $\sqrt{s_{\mathrm{NN}}} = 200$ GeV. Solid markers represent results from the HYDJET++ framework.}

\end{figure*}

\begin{figure*}[tbp]
\centering
\includegraphics[width=12cm]{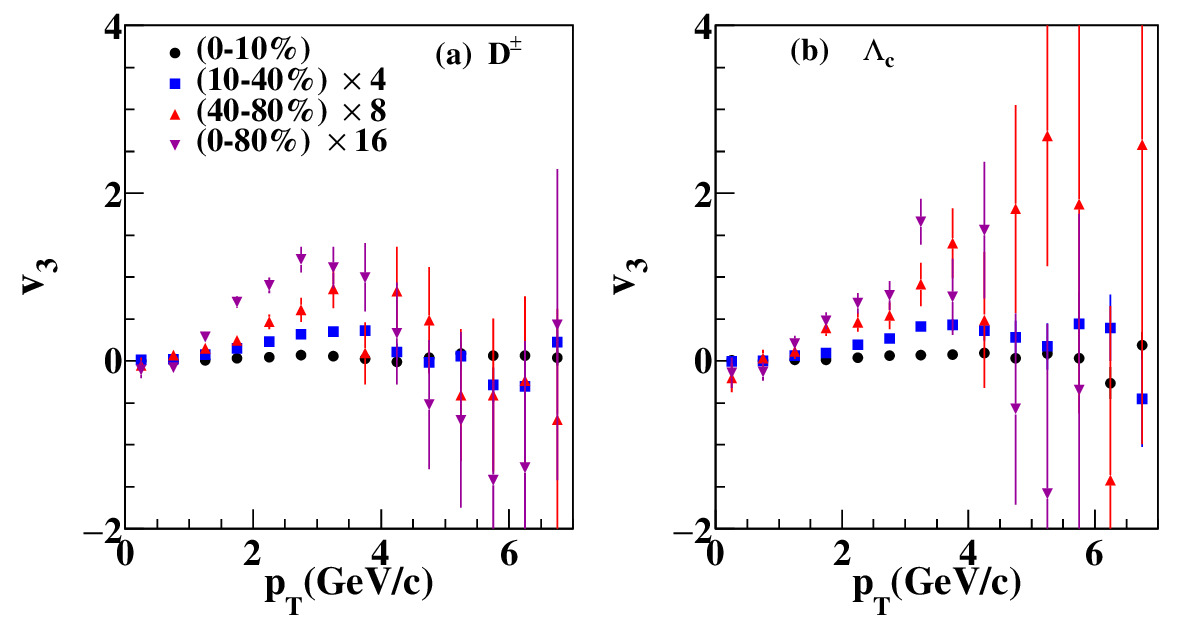}
\caption{\label{figure6} Triangular flow ($v_{3}$) of $D^{\pm}$ mesons and $\Lambda_{c}$ baryons as a function of transverse momentum ($p_{T}$) for various centrality bins in Au + Au collisions at $\sqrt{s_{\mathrm{NN}}} = 200$ GeV. Solid markers represent results from the HYDJET++ framework.}

\end{figure*}
Figure \ref{figure4}(a) represents the variation of elliptic flow ($v_{2}$) of the $D^{0}$ meson with $p_{T}$ for the 0-80\% centrality bin using the HYDJET++ model. Along with this, results from other models such as DUKE \cite{Cao:2015hia}, SUBATECH \cite{Ozvenchuk:2014rpa, Nahrgang:2014vza}, TAMU with charm quark diffusion (c-diff.) \cite{He:2011qa},  TAMU without charm quark diffusion (w/o c-diff.) \cite{He:2011qa}, and AMPT \cite{Lin:2004en} are also presented, which are extracted from these references \cite{Cao:2015hia, Ozvenchuk:2014rpa, Nahrgang:2014vza, He:2011qa, Lin:2004en}. These model predictions are compared with experimental results from the STAR experiment \cite{Liang}. It can be seen that the DUKE model describes the experimental data well up to $p_{T}=1$ GeV/c, while the SUBATECH model agrees with the data up to $p_{T}=2$ GeV/c. The TAMU model without charm quark diffusion is unable to describe the data well, whereas the TAMU model with charm quark diffusion matches up to  $p_{T}=2$ GeV/c. The AMPT model results are unable to describe the data at intermediate $p_{T}$ (2 GeV/c $<p_{T}<$ 4 GeV/c). Among all these models, our model provides the best description of the experimental data \cite{Liang} up to $p_{T}=4$ GeV/c. This highlights the significance of our model in describing the anisotropic flow compared to other models.

\begin{figure*}[tbp]
\centering
\includegraphics[width=15cm]{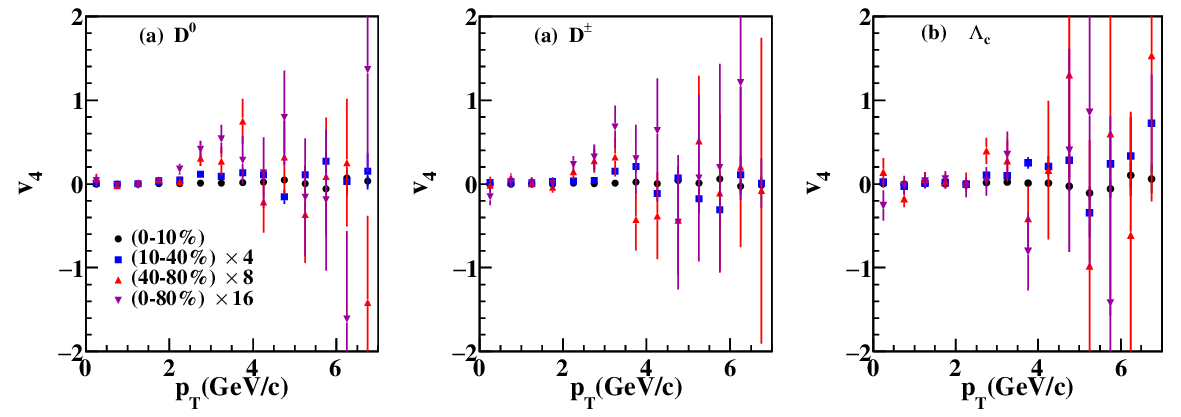}
\caption{\label{figure7} Quadrangular flow ($v_{4}$) of $D^{0}$ and $D^{\pm}$ mesons, and $\Lambda_{c}$ baryons as a function of transverse momentum ($p_{T}$) for various centrality bins in Au + Au collisions at $\sqrt{s_{\mathrm{NN}}} = 200$ GeV. Solid markers represent results from the HYDJET++ model.}

\end{figure*}

Figure \ref{figure4}(b) shows the triangular flow of the $D^{0}$ meson using the HYDJET++ model for various centrality bins. The SUBATECH model \cite{Ozvenchuk:2014rpa, Nahrgang:2014vza} results are also presented. As evident from Figure \ref{figure4}(a), the SUBATECH model describes the data up to $p_{T}=2$ GeV/c, which explains why its results match the HYDJET model results up to $p_{T}=2$ GeV/c in this case. In comparison, our model provides a better overall description of the results than the SUBATECH model.

\subsection{Predictions from HYDJET++ model}
\label{section3}
\subsubsection{$v_{n}$ spectra of heavy hadrons}

\begin{figure*}[tbp]
\centering
\includegraphics[width=15cm]{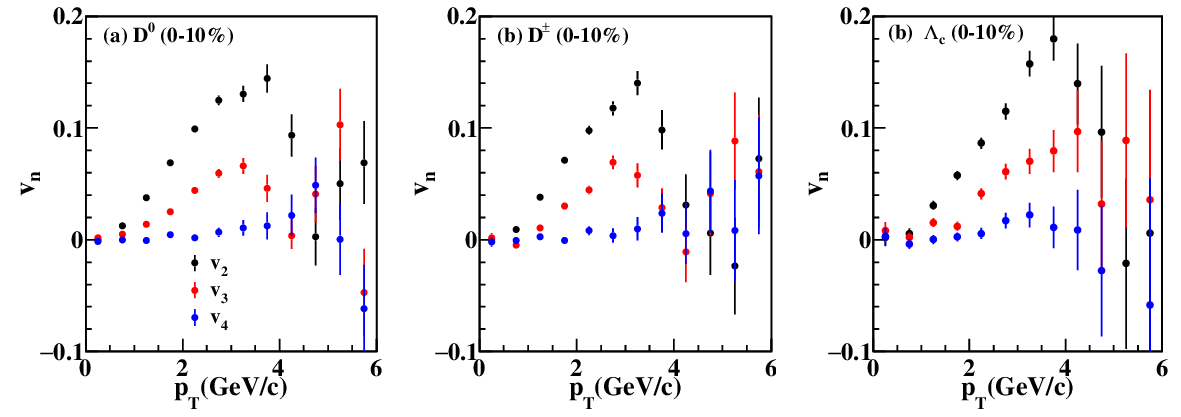}
\caption{\label{figure8} Anisotropic flow $v_{n}$ ($n = 2, 3, 4$) of $D^{0}$ and $D^{\pm}$ mesons, and $\Lambda_{c}$ baryons as a function of transverse momentum ($p_{T}$) for the 0–10\% centrality bin in Au + Au collisions at $\sqrt{s_{\mathrm{NN}}} = 200$ GeV. Solid markers represent results from the HYDJET++ model.}
\end{figure*}

\begin{figure*}[tbp]
\centering
\includegraphics[width=15cm]{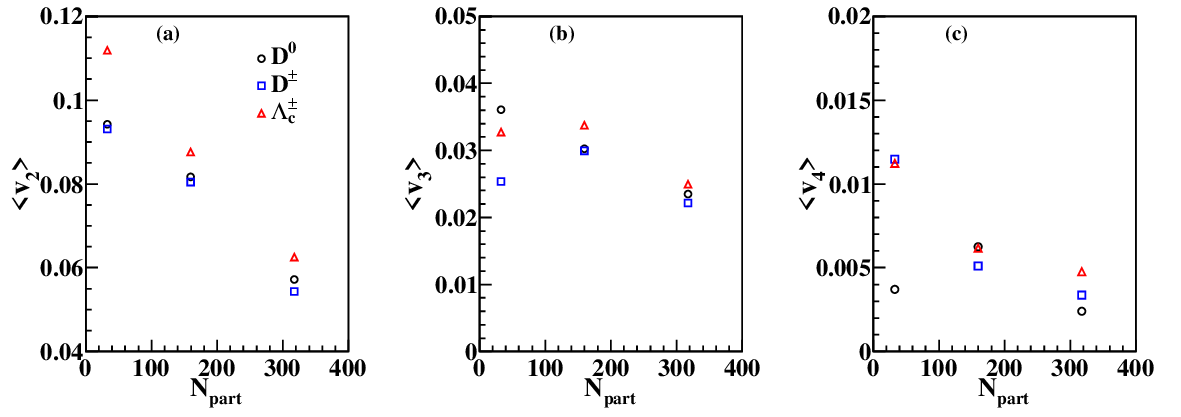}
\caption{\label{figure9}Average flow coefficients $\langle v_{2} \rangle$, $\langle v_{3} \rangle$, and $\langle v_{4} \rangle$ as a function of the number of participants in Au + Au collisions at $\sqrt{s_{\mathrm{NN}}} = 200$ GeV. Open markers represent results from the HYDJET++ model.}

\end{figure*}

Figure \ref{figure5} represents the predicted elliptic flow ($v_{2}$) of $D^{\pm}$ and $\Lambda_{c}$ as a function of $p_{T}$ for various centrality bins using the HYDJET++ model.  It is observed that, as we move from central to peripheral collisions, the peak of $v_{2}$ shifts to lower $p_{T}$. This occurs because the radial flow is smaller in peripheral collisions during the fireball expansion compared to central collisions. Our model results exhibit the same centrality dependence as observed in the experimental data \cite{Liang}. Figure \ref{figure6} represents the triangular flow of $D^{\pm}$ and $\Lambda_{c}$ as a function of $p_{T}$ for different centrality classes. It is observed that $v_{3}$ follows the same centrality dependence as $v_{2}$ in Figure \ref{figure5}. Figure \ref{figure7} represents the quadrangular flow ($v_{4}$) of $D^{0}$, $D^{\pm}$, and $\Lambda_{c}$ as a function of $p_{T}$ for various centrality classes. It can be seen that $v_{4}$ exhibits the same centrality dependence as $v_{2}$ in Figure \ref{figure5} and $v_{3}$ in Figure \ref{figure6}.

Figure \ref{figure8} represents the anisotropic flow ($v_{n}$) of heavy hadrons in Au + Au collisions at $\sqrt{s_{NN}}=200$ GeV. The magnitude of the triangular flow is observed to be smaller than that of the elliptic flow, and the quadrangular flow has an even smaller magnitude than the triangular flow. This indicates that the contribution of event-by-event fluctuations to the flow during the QGP expansion is relatively small.

Figure \ref{figure9} represents the average flow ($\langle v_{2} \rangle$, $\langle v_{3} \rangle$, and $\langle v_{4} \rangle$) of heavy hadrons as a function of the number of participants ($\langle N_{part} \rangle$) in Au + Au collision at $\sqrt{s_{NN}}=200$ GeV. It is observed that the integrated anisotropic flow decreases as $\langle N_{part} \rangle$ increases. The integrated anisotropic flow is mainly influenced by the initial geometric deformation and the energy loss within the QGP medium. For small $\langle N_{part} \rangle$, the geometrical anisotropy becomes larger while the energy loss effect is negligible \cite{Hong:2025dfj}. Additionally, a baryon–meson grouping is observed in $\langle v_{n} \rangle$, which is visible in the most central and mid-central collisions.

\begin{figure*}[tbp]
\centering
\includegraphics[width=15cm]{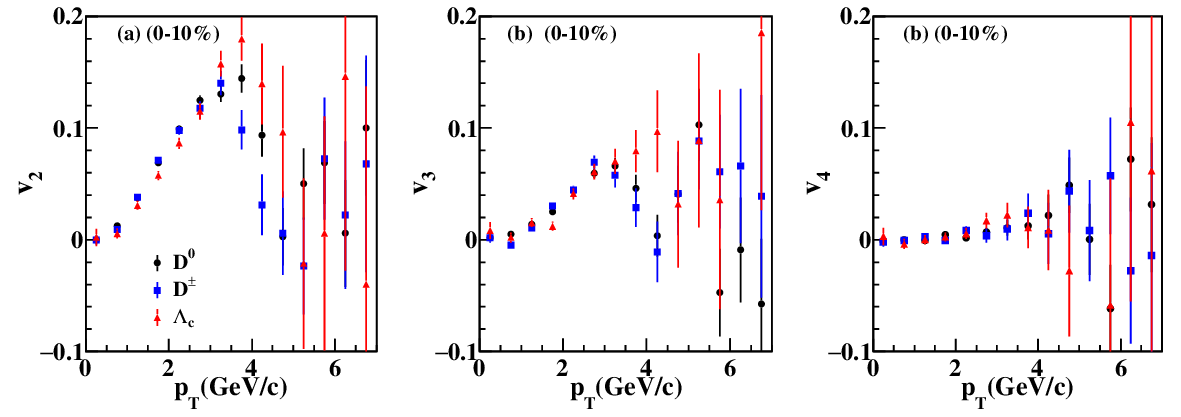}
\caption{\label{figure10} Anisotropic flow ($v_{n}$) of heavy hadrons as a function of transverse momentum ($p_{T}$) for the 0–10\% centrality bin in Au + Au collisions at $\sqrt{s_{\mathrm{NN}}} = 200$ GeV. Solid markers represent results from the HYDJET++ model.}

\end{figure*}

\begin{figure*}[tbp]
\centering
\includegraphics[width=15cm]{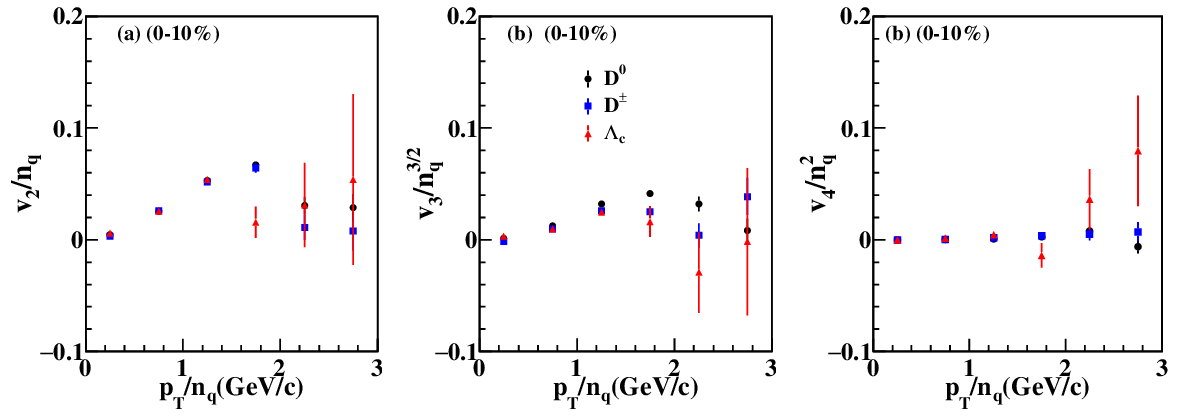}
\caption{\label{figure11} NCQ scaling of heavy hadrons for the 0–10\% centrality bin in Au + Au collisions at $\sqrt{s_{\mathrm{NN}}} = 200$ GeV. Solid markers represent results from the HYDJET++ model.}

\end{figure*}

\subsubsection{Mass ordering and NCQ scaling in heavy hadrons}
Figure \ref{figure10} represents the mass ordering in the flow coefficients $v_{2}$, $v_{3}$, and $v_{4}$ for heavy hadrons in Au + Au collision at $\sqrt{s_{NN}}=200$ GeV. At low $p_{T}$, a mass ordering among all heavy hadrons is observed in $v_{2}$ ($(v_{2})_{\Lambda_{c}}  < (v_{2})_{D^{\pm}}   < (v_{2})_{D^{0}}$), resulting from the combined effects of radial and elliptic flow. During the fireball expansion, radial flow pushes heavy hadrons to higher $p_{T}$. As a result, at low $p_{T}$,  the flow of baryons is smaller compared to that of mesons. At intermediate $p_{T}$ ($3<p_{T}<5$ GeV/c), the flow of baryons becomes larger than mesons due to the dominance of the coalescence mechanism of hadronization \cite{STAR:2003wqp}. In the cases of $v_{3}$ and $v_{4}$, the mass ordering is less pronounced.

  Figure \ref{figure11} represents the NCQ scaling of $v_{2}$, $v_{3}$, and $v_{4}$ for all heavy hadrons in the most central Au + Au collisions. The NCQ scaling reflects the dominance of partonic degrees of freedom in the early stages of the collision \cite{Jia:2006vj}. Some model calculations predict that the initial eccentricity ($\epsilon_{n}$) is proportion to $v_{n}^{2/n}$, implying that for the $n^{th}$-harmonic, the appropriate scaling factor should be $n_{q}^{n/2}$ instead of $n_{q}$ \cite{Lacey:2011av}. After applying this scaling, it is observed that $v_{2}/n_{q}$, $v_{3}/n_{q}^{3/2}$, and $v_{4}/n_{q}^{2}$ all lie on a universal curve for all hadrons up to $p_{T}=1.5$ GeV/c. This behavior suggests that charm hadrons originate from a deconfined, collectively expanding medium in which quarks are the fundamental interacting degrees of freedom. Thus, $v_{2}$, $v_{3}$, and $v_{4}$ exhibit NCQ scaling up to $p_{T}=1.5$ GeV/c.

\section{Summary and Outlook}
\label{sec:summ}

In this study, we investigated the anisotropic flow of heavy hadrons in Au + Au collisions at $\sqrt{s_{NN}}$ = 200 GeV using the HYDJET++ model. Our results for the elliptic ($v_{2}$), triangular ($v_{3}$), and quadrangular ($v_{4}$) flow coefficients of $D^{0}$, $D^{\pm}$, and $\Lambda_{c}$ hadrons align with experimental measurements from the STAR experiment, particularly at low and intermediate $p_{T}$ for central and mid-central collision. At high $p_{T}$, the discrepancy between the model results and experimental data is attributed to non-flow effects. In peripheral collisions, the discrepancy between model and experimental data at intermediate $p_{T}$ can be reduced by including mini-jet production. The model effectively reproduces several key features. These include the linear rise and subsequent fall of $v_{2}$ and $v_{3}$ with $p_{T}$, the centrality-dependent shift of the $v_{2}$ peak towards lower $p_{T}$, and the baryon–meson grouping observed in the average flow coefficients. Additionally, comparison with other theoretical models like DUKE, SUBATECH, TAMU, and AMPT demonstrates that HYDJET++ provides a more accurate description of heavy-flavor anisotropic flow up to $p_{T}=4$ GeV/c. 
Furthermore, the study confirms the presence of NCQ scaling in $v_{2}$, $v_{3}$, and $v_{4}$ for heavy hadrons, as well as mass ordering at low $p_{T}$, consistent with experimental observations. 

These findings reinforce the capability of the HYDJET++ model to effectively describe the collective behavior of heavy-flavor hadrons and enhance our understanding of the quark-gluon plasma dynamics in heavy-ion collisions. This study motivates us to investigate further the production of heavy quarkonia, such as $J/\psi$, $\psi'$, and $\chi_{c}$, which will be the focus of our future work.

\backmatter

\bmhead{Acknowledgements}

AK gratefully acknowledges financial support from the Institution of Eminence (IoE), Banaras Hindu University (BHU). SS and RG acknowledge financial support from the University Grants Commission (UGC) under the research fellowship scheme for central universities.

\bmhead{Data availability}
This manuscript is supported by experimental data, which are publicly available in the data repository referenced in \cite{STAR:2017kkh}. Additional data have been digitized from the theses referenced in \cite{Michael, Liang}.

\bmhead{Code availability} 

All data, code, and software from the HYDJET++ model used in this manuscript will not be deposited in a public repository but are available from the corresponding author upon reasonable request. The code and software generated and/or analyzed during the current study can also be obtained from the corresponding author upon reasonable request.

\sloppy
\bibliography{sn-article}

\end{document}